\documentclass[aps,pra,onecolumn,nofootinbib,superscriptaddress,showpacs]{revtex4}
\usepackage{amsmath,amssymb,amsthm}
\usepackage{amsfonts}
\usepackage{amssymb}
\usepackage{graphicx}
\usepackage{comment}
\usepackage{color}
\usepackage{natbib}
\usepackage{soul}

\usepackage[title]{appendix}

\begin{document}
	
\title{Information Disturbance Tradeoff in Bidirectional QKD}
	
\author{Nur Rahimah Sakinah Abdul Salam}
\affiliation{Faculty of Science, International Islamic University Malaysia (IIUM),
Jalan Sultan Ahmad Shah, Bandar Indera Mahkota, 25200 Kuantan, Pahang, Malaysia}
\affiliation{IIUM Photonics and Quantum Centre (IPQC), International Islamic University Malaysia (IIUM),
Jalan Sultan Ahmad Shah, Bandar Indera Mahkota, 25200 Kuantan, Pahang, Malaysia}
	
\author{Jesni Shamsul Shaari}
\affiliation{Faculty of Science, International Islamic University Malaysia (IIUM),
Jalan Sultan Ahmad Shah, Bandar Indera Mahkota, 25200 Kuantan, Pahang, Malaysia}
\affiliation{IIUM Photonics and Quantum Centre (IPQC), International Islamic University Malaysia (IIUM),
Jalan Sultan Ahmad Shah, Bandar Indera Mahkota, 25200 Kuantan, Pahang, Malaysia}

\author{Stefano Mancini}
\affiliation{School of Science \& Technology, University of Camerino, I-62032 Camerino, Italy}
\affiliation{INFN Sezione di Perugia, I-06123 Perugia, Italy}
	
\date{March 27, 2024}
	
\begin{abstract}
Making use of the Quantum Network formalism of \textit{Phys. Rev. A,} \textbf{82} (2010) 062305, we present the case for quantum networks with finite outcomes, more specifically one which could distinguish only between specific unitary operators in a given basis for operators. Despite its simplicity, we proceed to build a network derived from the optimal strategy in \textit{Phys. Rev. A,} \textbf{82} (2010) 062305 and show that the information-disturbance tradeoff in distinguishing between two operators acting on qubits, selected from mutually unbiased unitary bases is equal to the case of estimating an operator selected randomly from the set of SU($2$) based on the Haar measure. This suggests that such strategies in distinguishing between mutually unbiased operators is not any easier than estimating an operator derived from an infinite set. We then show how this network can be used as a natural attack strategy against a bidirectional quantum cryptographic protocol. 		
\end{abstract}

\pacs{-}

\maketitle
\section{Introduction}
\noindent
The basic tenets of quantum mechanics have introduced a profoundly new understanding of information when quantum systems are employed as carriers of information. The measurement postulate of quantum mechanics which notes the probability of perturbing a quantum system upon measurement unless the state is prepared as an eigenstate of the measured observable translates as the inability of gleaning information regarding a quantum degree of freedom without introduing disturbance or noise. The idea of information-disturbance in state estimation \cite{bib1} is closely connected with the notion of mutually unbiased basis (MUB) \cite{bib2}; i.e. measuring a state in a basis mutually unbiased to the one it is prepared in results in maximal uncertainty as well as introduces maximal disturbance (for any subsequent measurement made in the basis the state was originally prepared). This unique property proves invaluable in quantum cryptography, as it allows the detection of any intruders attempting to obtain information by measuring the quantum system, leaving behind a detectable trail of error or disturbance. This information-disturbance tradeoff has been successfully capitalised on in quantum cryptography, which is based on the notion of secure information shared between two communicating parties in the presence of an adversary.

Quantum cryptography, or more specifically Quantum Key Distribution (QKD), has evolved since its inception in 1984 \cite{bib3}. The first well-known protocol, BB84, utilizes four quantum states from two MUBs shared between legitimate communication parties, Alice and Bob, to establish shared secret keys. The protocol itself is extremely simple; Alice sends a large number of qubits to Bob over a quantum channel, and Bob measures the received qubits, randomly, in one of two measurement bases. A public discussion on the bases used (in preparation by Alice and measurement by Bob) after the end of the transmission and measurements of qubits would allow the parties to share a raw key. Any third party, commonly referred to as Eve, who would intercept the communication on the quantum channel, would induce errors whenever she taps into the quantum channel between Alice and Bob. Sampling for errors in the raw key would allow Alice and Bob to determine if a secure key can be established, where an error below a certain threshold would see the legitimate parties performing classical error correction procedure and privacy amplification to distill a secret key \cite{bib4,bib5}.
A different class of QKD protocols where encoding is done via unitary operators rather than preparation of qubits in different MUBs was later developed in the context of Two-Way QKD or Bidirectional Quantum Key Distribution (Bi-QKD) \cite{bib6}. The protocol involves a  bidirectional use of a quantum channel where a qubit prepared by Bob would be sent to Alice for some unitary operation randomly selected from a set of operations, and returned to Bob for his subsequent measurements. Bob then needs to infer Alice's operation by observing the evolution of the qubit. In short, any eavesdropping by Eve would require a proper estimation of unitary operators while inducing as little disturbance as possible to the system to evade detection. 

The issue of information-disturbance tradeoff in estimating an unknown unitary operator selected from a set of infinite number of such transformation, namely SU($2$), randomly distributed based on the Haar measure has been completely solved in \cite{bib7}. However, to date, there is no known formulation for a similar scenario given a finite set. While it has been conjectured that a more favorable scenario would emerge, the method and mathematics involved are expected to be non-trivial, as techniques and theorems applicable for the infinite case may no longer hold true. It is worth noting the obvious here; namely, the problem of a finite number of transformations is intimately connected to the security analysis of bidirectional QKDs, where a tight security proof, to date, is still lacking. 

In this paper, we explore the notion of quantum networks which provide only a finite set of possible results when testing any unitary operator. We nevertheless emphasize on distinguishing between 2 given unitary operators taken from some differing bases for operators. We then apply a relevant quantum network as an attack strategy against a bidirectional QKD protocol. Beginning with a quick review on the mathematical preliminaries and notation for constructing the quantum network in section w, we then introduce two quantum networks, which we refer to as the \textit{``projective network"} and the \textit{``optimal-I network"} in section 3. In general, such quantum networks would provide guesses of a unitary tested and subsequently, the whole system would behave as some unitary operator (not necessarily identical to the tested operator) acting on some input qubit. Our use of the term projective network is mainly motivated by how the network would effectively `project' the tested operator onto the subspace defined by the possible outcomes. Operationally, the effective operator that acts on an input quantum state would be the guessed unitary operator. This is akin to the projection of a quantum state onto some subspace defined by eigenstates of an observable basis for projective measurements. The term `optimal-I' reflects the fact that the network we construct is derived from the case in \cite{bib7}, corresponding to the optimal strategy to reduce disturbance while gaining as much information as possible. Here, we analyze, for each network, the probability of outcomes when testing or distinguishing between two unitary operators from two mutually unbiased bases, as well as the fidelity of the overall resulting channel. Furthermore, we compare the information tradeoff and disturbance of the optimal-I network with the approach presented in \cite{bib7} and we note a most interesting equivalence of our work in distinguishing between two specific operators to that of  \cite{bib7} in estimating a unitary operator selected from an infinite set. We further explore the application of a quantum network with Eve acting as an intruder in a bidirectional QKD protocol in section 4. Finally, we draw our conclusion in section 5.

\section{Preliminary and Notation}
	
\noindent
In this section, we provide a brief overview of the mathematical preleminary and notation used in the context of quantum networks \cite{bib7,bib8}.

We let $\mathcal{H}_{0},\mathcal{H}_{1},\dots,\mathcal{H}_{n}$ all be $d$-dimensional Hilbert spaces with subscripts denoting different systems. Except for the Hilbert spaces, calligraphic letters, like $\mathcal{M}$, will be used to denote maps mapping from one space to another. Operators on the other hand will be denoted by italicised capitals, like $M$. The space of linear operators from $\mathcal{H}_{a}$ to $\mathcal{H}_{b}$ is denoted as $\mathcal{L}(\mathcal{H}_a,\mathcal{H}_b)$ ($\mathcal{L}(\mathcal{H})$ whenever $\mathcal{H}_{a}=\mathcal{H}_{b}=\mathcal{H}$).

Throughout this work, we will also be working with (not necessarily normalized) maximally entangled bipartite states which are isomorphic to some unitary operator. For such purposes, we will make use of the double ket notation as follows; a unitary operator, $M$ in $\mathcal{L}(\mathcal{H}_a,\mathcal{H}_b)$ can be represented as  $M \equiv \sum_{i,j} \langle i | M |j \rangle|i\rangle |j \rangle = | M \rangle \! \rangle_{b,a}$, where $| M \rangle \! \rangle_{b,a} \in \mathcal{H}_b \otimes \mathcal{H}_a$. Here, the set of $\{|i\rangle\}$ and $\{|j \rangle\}$ represent fixed orthonormal bases for $\mathcal{H}_a$ and $\mathcal{H}_b$ respectively. The double-ket notation, $| M \rangle \! \rangle_{b,a}$, further satisfies the following property with regards to the unitary operator, $M$,
\begin{eqnarray}\label{eq1}
	| M \rangle \! \rangle_{b,a} = (M\otimes I) |I \rangle\! \rangle_{b,a} = (I \otimes M^T) |I\rangle\! \rangle_{b,a},
\end{eqnarray}
where $T$ represents transposition and $|I\rangle\! \rangle_{b,a} := \sum_{m=0}^{d-1} |m\rangle_{b} |m\rangle_{a}$. This isomorphism also gives the relationship for the inner products of the spaces as, $\langle\! \langle M| N \rangle\! \rangle =\text{Tr}[M^\dagger N]$ in $\mathcal{L}(\mathcal{H})$.  

We now move to defining quantum networks. A \textit{Quantum Network} consists of a concatenation of $N$ quantum channels with an initial input and some final output states. Operationally, quantum networks are meant to provide information of a given system tested by the network by comparing the input state and the output post-measurement states. More rigorously, a quantum network, denoted as $\mathcal{R}^{(N)}$, can be represented as a concatenation of $N$ linear maps, where $\mathcal{R}^{(N)}= \mathcal{W}_{1} \ast \mathcal{W}_{2} \ast \dots \ast \mathcal{W}_{N}$. Each linear map, referred to as $\mathcal{W}_{N}$, corresponds to a quantum channel. The network can be represented using \textit{wires} connecting linear maps in a circuit-like fashion as in figure 1. 
\begin{figure}[htbp]\label{Fig1}
	\centering % centering figure
	\scalebox{0.07} % rescale the figure by a factor of 0.8
	{\includegraphics{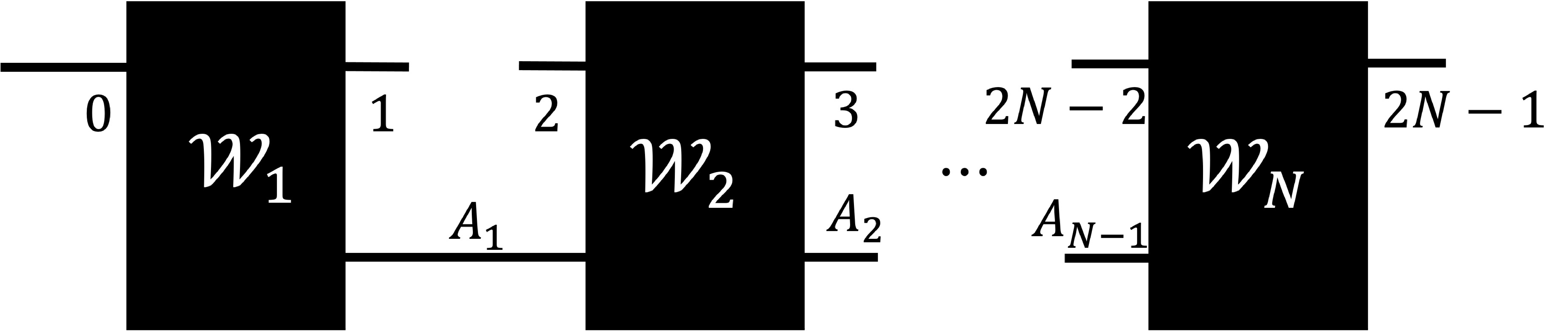}} % importing figure
	\caption{Structure of $N$ Quantum network, with quantum channel, $\mathcal{W}^{(1)}$ to $\mathcal{W}^{(N)}$  where $\mathcal{W}_j: \mathcal{L}(\mathcal{H}_{2j-2} \otimes \mathcal{H}_{A_{j-1}}) \rightarrow \mathcal{L}(\mathcal{H}_{2j-1} \otimes \mathcal{H}_{A_{j}})$.}
\end{figure}\\
The numbered wires correspond to input and output spaces of the differing Hilbert spaces and are labeled accordingly to represent the input and output spaces of the linear maps. The wires at the bottom with subscripts $A_{1},A_{2}, \dots, A_{N-1} $ in the circuit refer to the ancillary spaces in the quantum network and $\mathcal{H}_{A_0}:=\mathbb{C}$. The nature of the channels, either being trace preserving or otherwise\footnote{all channels must however be maps of non-trace increasing} is used to distinguish between two kinds of network. A \textit{deterministic} quantum network, is a concatenation of $N$  channels with all $\mathcal{W}_{N}$ being trace preserving maps. A \textit{probabilistic} quantum network on the other hand, is a concatenation of $N$ linear maps with at least one map being trace decreasing. 

In what follows, we will make use of Choi operators for a map and it is instructive to digress briefly for a quick summary on the matter. Beginning with a map, say, $\mathcal{Q}$, defined as $\mathcal{Q}$ :$\mathcal{L}(\mathcal{L}(\mathcal{H}_a),\mathcal{L}(\mathcal{H}_b))\rightarrow\mathcal{L}(\mathcal{H}_a \otimes \mathcal{H}_b)$, the Choi isomorphisim \cite{bib8} gives a corresponding linear operator $Q \in \mathcal{L}(\mathcal{H}_a \otimes \mathcal{H}_b)$ as,
\begin{eqnarray}\label{eq2}
	Q := (\mathcal{Q} \otimes \mathcal{I}_a)(|I\rangle \! \rangle \langle \! \langle I |_{a,a}),
\end{eqnarray}
where $\mathcal{I}_a$ is the identity map on $\mathcal{L}(\mathcal{H}_a)$. The linear operator $Q$ is called the Choi operator of the linear map $\mathcal{Q}$. 

A \textit{Generalized $N$-instrument} is defined as a set of probabilistic quantum network $\{\mathcal{R}_j\}$ whose sum is a deterministic quantum network $\mathcal{R}^{(N)}= \sum_{j} \mathcal{R}_j$ and satisfy the normalization condition given by:
\begin{eqnarray}\label{eq3}
	\text{Tr}_{2p-1}[R^{(p)}] =  I_{2p-2} \otimes R^{(p-1)} , ~ 1 \leq p \leq N.
\end{eqnarray}
The subscript for the identity operator denotes the index for the Hilbert space it acts on. Thus, we can write, the Choi operator $R^{(N)} \in \mathcal{L}(\mathcal{H}_{i} \otimes \mathcal{H}_{o}) $  where the input space $\mathcal{H}_{i} = \otimes^{N}_{q=1} \mathcal{H}_{2q-2} \rightarrow \mathcal{H}_{i_{p}} = \otimes^{p-1} _{q=0} \mathcal{H}_{2q}$ and the output space $\mathcal{H}_{o} = \otimes^{N}_{q=1} \mathcal{H}_{2q-1}  \rightarrow \mathcal{H}_{o_{p}} = \otimes^{p-1} _{q=0} \mathcal{H}_{2q+1}$ and $R^{(0)}=1$. The normalization condition emphasizes the causal ordering in the deterministic quantum channel.

Now let us consider an example of a quantum network, $\mathcal{R}^{(N)}$, from a Hilbert space, $\mathcal{H}_1$ to $\mathcal{H}_2$,  and a quantum network, $\mathcal{Q}^{(M)}$, from $\mathcal{H}_0$ to $\mathcal{H}_1$, where both quantum networks can be \textit{linked} together as $\mathcal{R}^{(N)} \star \mathcal{Q}^{(M)}$. The \textit{link product} between two quantum networks can be written in terms of their Choi operators as,
\begin{eqnarray}\label{eq4}
	R^{(N)} \ast Q^{(M)} = \text{Tr}_{1} \Big[ ( R^{(N)}_{2,1} \otimes I_{0}) ( I_{2} \otimes Q^{(M)T_{1}}_{1,0} )\Big],
\end{eqnarray}
where $R^{(N)}, Q^{(M)}$ are the Choi operators of $\mathcal{R}^{(N)}$ and $\mathcal{Q}^{(M)}$ respectively. The symbol $T_{1}$ here denotes the partial transposition with respect to fixed orthonormal basis in Hilbert space $\mathcal{H}_1$.

In estimating a unitary operator, we choose a quantum network with $N=2$, described as a Quantum 2-Tester \cite{bib7,bib8}; i.e. a sequence of two quantum channels with input and output spaces and an empty slot. Figure 2 illustrates the structure of a unitary channel, $\mathcal{P}$, \textit{plugged into} the empty slot of the quantum network.\\
\begin{figure}[htbp]\label{Fig2}
	\centering % centering figure
	\scalebox{0.06} % rescale the figure by a factor of 0.8
	{\includegraphics{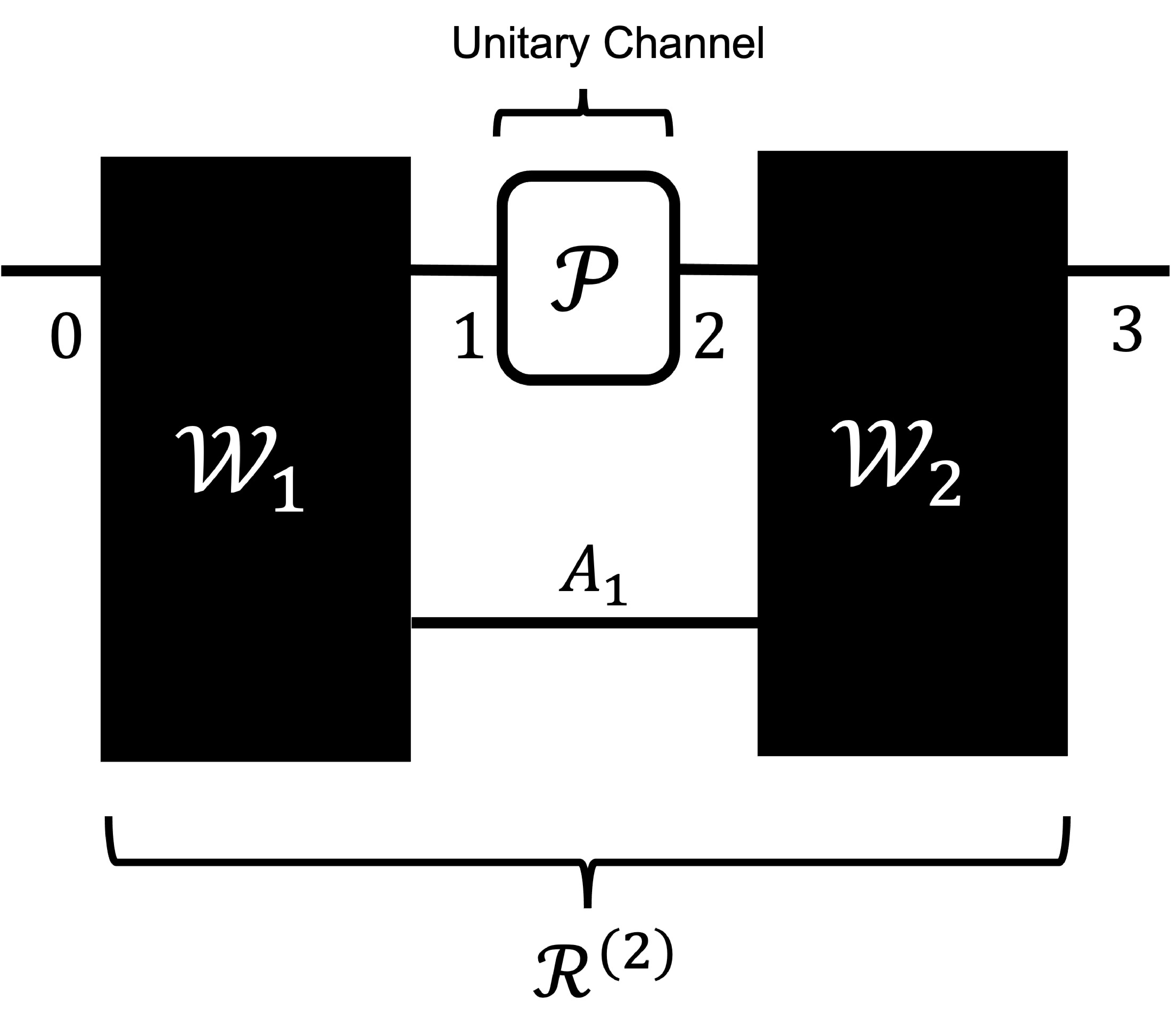}} % importing figure
	\caption{Structure of Quantum network for $N=2$, linked to the unitary channel, $\mathcal{P}$.}
\end{figure}\\
The normalization and generalized instruments from \ref{eq3} for the Quantum 2-Tester $ R^{(2)} := \sum_{i=1}^{d^2 -1} R_i$ \cite{bib7} is given as follows:
\begin{eqnarray}\label{eq5}
	\text{Tr}_{3}[R^{(2)}] = & ~ I_{2} \otimes R^{(1)} ~~,~~\text{Tr}_{1}[R^{(1)}] =  ~ I_{0}.
\end{eqnarray}

In our work, we will focus on quantum networks which provide an outcome, $i$, which is a guess for the unitary tested as $U_i$ from a finite set of possible outcomes\footnote{the index also numbers the elements in the finite set} when estimating some unitary operator, $P$. Particularly, we are considering the unitary operator, $U_i$, chosen from a finite basis, for a $d^2$-dimensional space of operators. The probability of such an outcome, i.e the probability of having a guess, $U_i$, given the operator, $P$, is,
\begin{eqnarray}\label{eq6}
	\text{Pr}(U_{i}|P) = \frac{1}{d} \text{Tr}[R_{i} (I_3 \otimes |P^* \rangle\!\rangle \langle\!\langle P^*| _{2,1} \otimes I_0)],
\end{eqnarray}
where the operator $|P^* \rangle\!\rangle \langle\!\langle P^* |_{2,1}$ is written in double ket notation in the Hilbert space $\mathcal{H}_2 \otimes \mathcal{H}_1$.
The information gain for a specific outcome can be quantified using the following formula \cite{bib7}:
\begin{eqnarray}\label{eq7}
	g(U_i,P) = \frac{1}{d^2} |\text{Tr}[U_i P^\dagger]|^2,
\end{eqnarray}
which really is proportional to the inner product between the guessed unitary and the actual one. It's obvious to note that the gain achieves its maximum value of 1 only when the guess, $U_i$, is identical to $P$. The information gain, $G_{P}$, for the network given $P$ can be calculated as a weighted average based on the outcomes of the quantum network $\mathcal{R}^{(2)}$ in estimating the unitary operator $P$,
\begin{eqnarray}\label{eq8}
	G_{P}= \sum_{i=1}^{d^2} \Pr(U_i |P) g(U_i ,P).
\end{eqnarray}
The summation from $i=1$ to $d^2$ reflects the elements in a finite basis for the operator space. The performance of the overall quantum operation $R^{(2)} * P$ in comparison to $P$ can be quantified in terms of the fidelity given by \cite{bib7} as,
\begin{align}
	F(R^{(2)}|P)  = &  \frac{1}{d^2} \langle \! \langle P|_{3,0} \langle \! \langle P^{*}|_{2,1} R^{(2)}| P \rangle \! \rangle_{3,0} |P^{*}\rangle \! \rangle_{2,1} \label{eq9}\\
	 = &   \frac{1}{d^2} \sum_{i=1}^{d^2}\langle \! \langle P|_{3,0} \langle \! \langle P^{*}|_{2,1} R_i | P \rangle \! \rangle_{3,0} |P^{*}\rangle \! \rangle_{2,1}. \label{eq10}
\end{align}
We will make use of these quantities later on when assessing a tradeoff between information gain and disturbance in distinguishing two nonorthogonal unitary operators.

\section{Quantum Network in SU($2$)}
\noindent
We propose two specific networks using $N=2$ quantum channels, similar to those in \cite{bib7}. The first is a network that would estimate a unitary $P$ resulting in a guess $U_i\in\mathbb{U}$ with $\mathbb{U}$ being some unitary basis  and the network itself would subsequently behave as $U_i$. This may lead to possible disturbance if $U_i \neq P$. Akin to the notion of `projective measurements' in quantum state estimations, we refer to this network as ``\textit{projective network}". In a bid to minimize the disturbances of such a network, we introduce another quantum network called the ``\textit{optimal-i network}". This network allows us to have a handle on the disturbance at the expense of some information gain. We will then establish an information-disturbance tradeoff when employing such a network in terms of the earlier described figures of merit for information gain and fidelity.

\subsection{Projective Network in SU($2$)}

\noindent
Consider an orthogonal basis consisting of unitary operators, $\mathbb{U}=\{U_i\}$, for the $2^2$ dimensional space, $\mathcal{L}(\mathcal{H})$. The normalization condition for the generalized instrument, $\{ R_{i}\}$, for a projective network, $\mathcal{R}^{(2)} _\mathbb{U}$ for $\mathbb{U}$, is given as,
\begin{eqnarray}\label{eq11}
	R^{(2)}_{\mathbb{U}} = \sum_{i=1}^{4} R_{i}  ~,~ R_{i} = \frac{1}{4} ~ |U_{i}  \rangle \! \rangle \langle \! \langle U_{i}   | _{3,0}  \otimes  |U_{i}  ^* \rangle \!\rangle \langle\!\langle U_{i}  ^* | _{2,1},
\end{eqnarray}
and $R_i$ is a positive operator for all $i$. This deterministic quantum network fulfils the normalization conditions of \ref{eq5} and can be demonstrated as follows,
\begin{eqnarray}\label{eq12}
	\text{Tr}_3 [R^{(2)}_{\mathbb{U}}]   =  \frac{1}{4} ( I_0 \otimes  \sum_{i=1}^{4}  |U_{i}  ^* \rangle \! \rangle \langle \! \langle U_{i}  ^* |_{2,1}) = I_2 \otimes  R^{(1)}_{\mathbb{U}},
\end{eqnarray}
where $R^{(1)}_{\mathbb{U}} \in \mathcal{L} (\mathcal{H}_1 \otimes \mathcal{H}_0)$, with $\sum_{i=1}^{4}  |U_{i}  ^* \rangle \! \rangle \langle \! \langle U_{i}  ^* |_{2,1} = 2 (I_2 \otimes I_1)$  and $R^{(1)}_{\mathbb{U}} =(I_1 \otimes I_0)/2$, giving the last trace, $\text{Tr}_1 [ R^{(1)}_{\mathbb{U}}] = I_0$ with $R^{(0)}_{\mathbb{U}} = I_0$. The normalization condition for $R^{(2)}_{\mathbb{U}}  $ becomes trivial with $\text{Tr}[R^{(2)}_{\mathbb{U}} ] = 4$. 

In this work, unless otherwise stated, we shall consider $\mathbb{U}=\{U_i\}$ as the standard basis for $\mathcal{L}(\mathcal{H})$; i.e. comprising of the identity and the Pauli operators, more explicitly, $U_1=I,U_2=\sigma_x,U_3=\sigma_y$ and $U_4=\sigma_z$. There is no loss of generality in choosing this basis, as there are no reasons to assume that the standard basis can in some ways be preferred over any other.

\subsubsection{Distinguishing Between Two Unitary Operators}

\noindent
We begin by referring to figure \ref{Fig2} and consider  an arbitrary unitary operator $P=\sum_{j=1}^{4} a_{j}U_{j}$ in SU($2$) expressed in terms of the operators in $\mathbb{U}$ with $\sum_{j=1}^{d^2}|a_{j}|^2=1$. 
When estimating the operator, $P$, the probability of the outcome $U_{i} \in \mathbb{U}$ is $\Pr(U_{i}|P)=|a_{i}|^2$, which is also the information gain for the outcome, $i$, as defined in \ref{eq7}. Thus, the information gain, $G_{P}^{{Proj}}$, and fidelity, $F_{P}^{Proj}$, for the projective network in estimating an arbitrary unitary operator, $P$ are given by \ref{eq8} and \ref{eq9} respectively as,
\begin{eqnarray}\label{eq13}
	G_{P}^{{Proj}} = \sum_{i=1}^{4} |a_i|^4~,~F_{P}^{Proj} = \sum_{i=1}^{4} |a_i|^4.
\end{eqnarray}
With $1/4\leq\sum_{i=1}^{d^2} |a_i|^4 \leq 1$, it is not difficult to identify the cases where the extremes of the inequality is met. The maximum value of the gain (and fidelity of the channel) happens when $U_{i}=P$. On the other hand, the minimum can be achieved when $|a_i|^2=1/4$ for all $i$, or, $|\text{Tr}[U_i P^\dagger]|^2=1/4$, i.e. when the operators $U_i$ and $P$ are mutually unbiased. Thus, it would be interesting to understand the average gain and fidelity (and by extension the disturbance) when the projective network is used to distinguish between two unitary operators that are mutually unbiased with respect to one another. 

Let us then consider an operator to be selected randomly from the set $\{P_1, P_2\}$ with equal probability, where $P_1\in \mathbb{U}$, while $P_2\in  \mathbb{M} = \Big\{(U_1 + \sum_{k=2}^{4} \alpha_k U_k)/2 | \alpha_k \in \{\pm{1}\}, \prod_{k=2}^{4} \alpha_k = (-1) \Big\}$ is a mutually unbiased unitary basis, (MUUB) with respect to the standard basis, $\mathbb{U}$. Averaging over the use of the two operators, $P_1$ and $P_2$, with equal probability, the average information gain when distinguishing between the two can be obtained as 
\begin{eqnarray}\label{eq14}
	G_{avg}^{Proj^{\prime}} = \frac{1}{2} \sum_{i=1}^{4} \sum_{j=1}^{2} \Pr(U_i |P_j) g(U_i, P_j) = \frac{5}{8},
\end{eqnarray}
and the average fidelity as 
\begin{eqnarray}\label{eq15}
	F_{avg}^{Proj \prime} = \frac{1}{2} \sum_{i=1}^{d^2} \sum_{j=1}^{2} F(U_i |P_j) = \frac{5}{8}.
\end{eqnarray}
These values basically imply that, not only there is no way one can achieve an average gain of 1, but it is also impossible to have a fidelity of 1, or evade any perturbation to the channel when distinguishing nonorthogonal operators. It would be however possible to consider reducing the perturbation to the system at the cost of information gain; this is the motivation for the ensuing subsection.

\subsection{Optimal I-Network in SU($2$)}

\noindent
Similar to optimal network in estimating a unitary operator selected from a set of infinite possible operators \cite{bib7}, we consider the case for one that would distinguish only between two operators. Such a network would allow us to manage the disturbance caused when estimating an operator. We should note that, while \cite{bib7,bib8} noted the optimality of the network, it is not immediately obvious if our adaptation for the finite case is also optimal. However, we argue for its possible optimality in the sense that it would, in the extreme cases, desirably result in information gains (versus disturbances) identical to that of projective measurements. Let us consider the `optimal-I network' for the standard basis, $T^{(2)}_{\mathbb{U}}$, which can be expressed as a generalized instrument, $\{T_{i}\}$ given by $T^{(2)}_{\mathbb{U}} = \sum_{i=1}^{4} T_{i}$ , where 
\begin{align}
	T_{i}  & = \frac{1}{4} |  \mathcal{X}_{i} \rangle \! \rangle\langle \! \langle \mathcal{X}_{i} | , \label{eq16}\\	
	| \mathcal{X}_{i} \rangle \! \rangle & = x~| U_{i} \rangle \! \rangle _{3,0} | U_{i} ^* \rangle \! \rangle_{2,1} + y~|I \rangle \! \rangle_{3,2} |I \rangle \!\rangle_{1,0} ~,~ x,y \in \mathbb{R}^{+}. \label{eq17}
\end{align}
Note that the indices `$i$' of $T_i$ corresponds to $U_{i}\in\mathbb{U}$ and $x,y$ are real numbers such that $x^2+ xy + y^2=1,0 \leq x\leq 1,0 \leq y\leq 1$ \cite{bib7}. The normalization of the optimal-I network is shown as follows:
\begin{align}
	\text{Tr}_3 [T^{(2)}_{\mathbb{U}}]  & = \frac{1}{4} \Big[x^2  \Big(I_0 \otimes \sum_{i=1}^{4} |U_{i} ^{*}\rangle\!\rangle\langle\!\langle U_{i}^{*}|_{2,1} \Big) + 4xy\Big( I_{2} \otimes | I \rangle\! \rangle \langle \! \langle I |_{1,0} \Big) \nonumber \\
	& \quad+ 4y^2 \Big( I_2 \otimes | I \rangle\!\rangle \langle\!\langle I|_{1,0}    \Big) \Big] \label{eq18} \\
	& =  I_{2} \otimes T^{(1)}_{\mathbb{U}}, \label{eq19}
\end{align}
where $\sum_{i=1}^{4} |U_{i} ^{*}\rangle\!\rangle\langle\!\langle U_{i}^{*}|_{2,1} =2( I_2 \otimes I_1)$, and $T^{(1)}_{\mathbb{U}} = x^2(I_1 \otimes I_0)/2 +xy | I \rangle\!\rangle \langle\!\langle I |_{1,0} + y^2| I \rangle \! \rangle \langle \! \langle I |_{1,0}$.
The last trace of the normalization condition is given by:
\begin{align}
	\text{Tr}_1 [T^{(1)}_{\mathbb{U}}]  & = \text{Tr}_1 \Big[\frac{1}{2} x^2  (I_1 \otimes I_0) + xy | I \rangle \! \rangle \langle \! \langle I |_{1,0} + y^2| I \rangle \! \rangle \langle \! \langle I |_{1,0}\Big] \label{eq20}\\
	& =  (x^2+ xy + y^2) I_0, \label{eq21}
\end{align}
with $T^{(0)}_{\mathbb{U}} =  I_0$. The normalization condition for $T_{\mathbb{U}} ^{(2)}$ becomes trivial, as $\text{Tr}[T_{\mathbb{U}} ^{(2)} ] = 4 x^2 + 4xy + 4 y^2$, which is also equal to $4$. Similar to \cite{bib7,bib8}, we can observe that a maximal value of $x = 1$ implies a projective network, while the other extreme, $y = 1$, implies a passive operation by the identity network, meaning it does not induce any errors while also not gaining any information.

\subsubsection{Distinguishing Between Two Unitary Operators}

\noindent
Using the same approach as in the previous section in distinguishing between the operators $P_1$ and $P_2$, the probability of guessing for optimal-I network, $U_{i}^{T_{i} }$ given $P_1$ is 
 \begin{align}
	\text{Pr}(U_{i}^{T_{i}}|P_1) & = \begin{cases}
		x^2+xy+y^2/4  ~~\mbox{if}~~ U_i^{T_{i}}=P_1
		\\
		y^2/4 \qquad\qquad~~~~\mbox{if}~~ U_i^{T_{i}} \neq P_1
	\end{cases},\label{eq22} \\
	\text{Pr}(U_{i}^{T_{i}}|P_2)  & = \frac{1}{4}(x^2+xy+y^2) ~,~ \forall i. \label{eq23}
\end{align}
This is in contrast with the projective case, where the probability is zero to result in a guess for $U_i \neq P_1$. This is really where the limitation in information gain becomes apparent when compared to the projective network.

The average information gain, $G_{avg}^{Opt}$, when distinguishing between $P_1$ and $P_2$, is then given by
\begin{eqnarray}\label{eq24}
	G_{avg}^{Opt}= \frac{1}{2} \sum_{i=1}^{4} \sum_{j=1}^{2} \Pr(U_i^{T_{i}} |P_j) g(U_i^{T_{i}}, P_j) = \frac{5}{8} x^2+\frac{5}{8}xy +\frac{1}{4}y^2,
\end{eqnarray}
while the average fidelity, $F_{avg}^{Opt}$, is as follows,
\begin{eqnarray}\label{eq25}
	F_{avg}^{Opt} = \frac{1}{2} \sum_{i=1}^{2} [F(U_i^{T_{i}} ,P_1) + F(U_i^{T_{i}},P_2)] = \frac{5}{8}x^2 + xy + y^2.
\end{eqnarray} 
It is obvious to note that we regain the projective network when we set $x=1,(y=0)$. On the other hand, setting $y=1,(x=0$) would imply a passive operation in trying to glean information, i.e. not perturbing the original unitary channel in anyway and results in maximal fidelity while having no information gain.

\subsubsection{Information and Disturbance Tradeoff in Optimal-I Network}

\noindent
In assessing a tradeoff between information gain and disturbance caused by the optimal-I network, we use the quantification of information, $I$, and disturbance, $D$, \cite{bib7} expressed as,
\begin{equation}\label{eq26}
	I:= \frac{G_{avg}- G_{min}}{G_{max}-G_{min}} ~,~ D:= \frac{F_{max}-F_{avg}}{F_{max}-F_{min}} ,
\end{equation}
where $G_{avg}$ is the average information gain with $G_{min}$ and $G_{max}$ representing the minimum and maximum values respectively, and $F_{avg}$ is the average fidelity with $F_{min}$ and $F_{max}$ represent the minimum and maximum values respectively.
Making use of %equations 
\ref{eq24} and \ref{eq25}, the information and disturbance, $I_{Opt}$ and $D_{Opt}$ for the optimal-I network in distinguishing between $P_1$ and $P_2$ are thus given as,
\begin{equation}\label{eq27}
	I_{Opt} =  \frac{(\frac{5}{8}x^2+\frac{5}{8}xy+\frac{1}{4}y^2)-\frac{1}{4}}{\frac{5}{8}-\frac{1}{4}} ~,~
	D_{Opt} =  \frac{1-(\frac{5}{8}x^2+xy+y^2)}{1-\frac{5}{8}}=x^2.
\end{equation}
These parameters' relation to $x$ and $y$ are identical to that of \cite{bib7} despite the latter using a network to estimate an arbitrary unitary operator with the averages taken over the entire possible unitary choices in SU($2$). The normalization condition of $x^2+xy+y^2=1$ thus can be rewritten as
\begin{equation}\label{eq28}
	(D_{Opt}-I_{Opt})^2-D_{Opt}(1-I_{Opt})=0.
\end{equation}
Given the normalization condition above, we may write $x= (-y + \sqrt{4 - 3y^{2}})/2$ and we may then plot a parametric curve for $D_{Opt}$ versus $I_{Opt}$ for varying values of $y$ (and by extension, $x$). The plot is in fact the same as \cite{bib7} for SU($2$), which is certainly very interesting as it highlights a simple point; on average, distinguishing between 2 unitary operators selected from differing MUUBs is equivalent to estimating an operator selected randomly from SU($2$).    
\begin{figure}[htbp]\label{Fig3}
	\centering
	\includegraphics[width=0.8\linewidth]{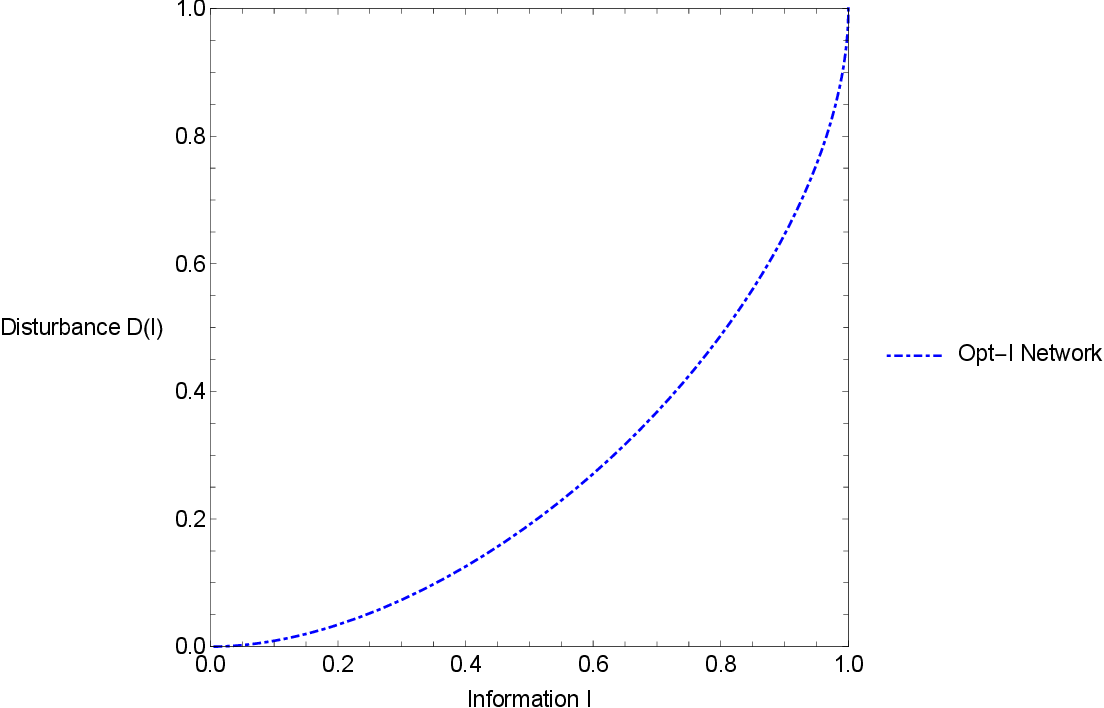}\\
	\caption{Information and Disturbance curve of optimal-I network.}
	%\label{fignorm}
\end{figure}

It is noteworthy that both the projective and optimal-I networks can be constructed for the MUUB, $\mathbb{M}$, instead of the standard one as we considered in our demonstrations above. It can be shown that such a network, $T^{(2)}_\mathbb{M}$ would give the same results for the information-disturbance curve. 

It is perhaps important to point out that despite our specific use of the standard basis $\mathbb{U}$, in our demonstrations, we do not suspect any loss of generality as the nature of mutually unbiased unitary bases relies on the relationship between two bases, a concept not exclusive to the standard basis. Coupled with the fact that mathematically speaking, no unitary operator basis is essentially unique, it’s safe to presume that the figures of merits relevant to estimation of unitary operators derived from any given bases should be independent of basis choice, i.e. one should always be able to construct some network given some basis to yield the same information-disturbance curves.

In the next section, we will explore the use of the optimal-I network in a bidirectional QKD protocol. Consequently, the quantum network can be regarded as a measurement apparatus to estimate unknown unitary operators which play the role of information encoder in such protocols.

\section{Bidirectional QKD}

\noindent
We consider a protocol similar to \cite{bib9}, which makes use of the quantum channel twice, i.e. Bob sends a qubit of his choice to Alice who would encode by executing a unitary transformation on the qubit before returning it to Bob for relevant measurements to determine Alice's encoding. 

Let Alice's unitary operator be either $A_{1}= U_1$ or $A_{2}=(U_1 +i U_2+i U_3- i U_4)/2$; i.e. operators randomly selected from two differing MUUBs \cite{bib10}, with equal probability. Bob prepares an input state $|\phi \rangle$ from either the computational basis, $\{|0\rangle,|1\rangle\}$ or the X basis, $\{ |X_{\pm}\rangle = (|0\rangle \pm |1\rangle)/\sqrt{2} \}$. Alice's unitary operator $A_1$ is a passive operation while $A_2$ corresponds to the flipping of prepared input states to one that is mutually unbiased to it as demonstrated in the following for the various input states that Bob may send
\begin{eqnarray}\label{eq29}
	A_{2}|0\rangle\rightarrow |X_{-}\rangle ~,~ A_{2}|1\rangle\rightarrow |X_{+}\rangle ~,~ A_{2}|X_+\rangle\rightarrow |Y_{+}\rangle ~,~ A_{2}|X_-\rangle\rightarrow |Y_{-}\rangle.
\end{eqnarray}  
Upon executing her chosen operation, Alice would forward the qubit back to Bob who will commit to random measurements in either $Z$ or $X$ basis for states he originally prepared in the $Z$ basis, or $X$ or $Y$ basis for states he originally prepared in the $X$ basis. The decoding by Bob is done through inferring what operator Alice \textit{could not} have used. This is basically similar to the decoding process of \cite{bib9}. If Bob had prepared an input state $|\phi\rangle$ and measured in the same basis, a resulting state orthogonal to $|\phi\rangle$ would inform Bob that Alice could not have used $A_1$ and thus conclusively infer that Alice has used $A_2$. Measurement in a different basis ($X$ if $|\phi\rangle$ was prepared in $Z$ and $Y$ otherwise) resulting in a state orthogonal to $A_2|\phi\rangle$ would allow Bob to conclusively infer that Alice has used $A_1$ instead.

Discarding all inconclusive results, it is possible for Alice and Bob to share a common key string by using the logical `0' for $A_1$ and `1' for $A_2$. We summarise the protocol as follows:
\begin{enumerate}
	\item 	Bob would randomly prepare a state selected from either the computational basis, $\{|0\rangle,|1\rangle\} $ or the X basis, $\{|X_{\pm}  \rangle= (|0\rangle \pm  |1 \rangle) / \sqrt{2}\}$ and submit to Alice.
	\item Alice would encode the qubit from Bob with either one of two operators $A_{1}$ and $A_{2}$ (as described above), before returning it back to Bob.
	\item Bob would perform a measurement in either $Z$ or $X$ basis ($X$ or $Y$) for states he originally prepared in the $Z$ ($X$) basis.
	%\item Bob sends input state in X basis, the measurement would be randomly choosen in X or Y basis
	\item After repeating steps 1-3 for a large number of times, Bob would note the cases where his measurements would allow him to infer Alice's encoding with certainty and discard the rest. He would subsequently share this information over the public channel with Alice and a raw key is established  based on the non-discarded measurement results only.
\end{enumerate}
The above in principle would allow perfect correlation between Alice and Bob. Realistically however, errors would be present and in establishing a secret key, such errors must be attributed to the eavesdropping attempts of an adversary, Eve. A standard estimation of error rates and an error correction procedure then need to be executed and in the event the error rate does not exceed some prescribed security threshold, a privacy amplification procedure is carried out to distill secret key for which an eavesdropper would have arbitrary small information of.

\begin{figure}[hbt!]\label{Fig4}
	\centering % centering figure
	\scalebox{0.65} % rescale the figure by a factor of 0.8
	{\includegraphics{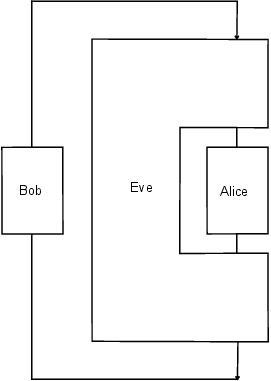}} % importing figure
	\caption{Schematic description of Eve's attack on bidirectional QKD protocol by means of quantum neteworks.}
	%\label{fig:norm} % labeling to refer it inside the text
\end{figure}

An obvious difference between protocols like this where Alice makes use of  nonorthogonal unitary operators versus those making use of orthogonal ones like \cite{bib11,bib9} would be the need of the Control Mode (CM) which is executed randomly by Alice. In such a mode, Alice would make measurements instead of encoding, to determine if there are any errors in a subsequent public discussion. This mode is crucial as Eve can commit to attack strategies that would allow her to distinguish perfectly between the operators used in encoding. Protocols like the above where Alice's operators are nonorthogonal in principle does not necessitate such a CM given the indistinguishability of such operators for a single use while introducing errors in the encoding process. The security analysis here thus becomes straightforward; we consider Eve committing to an attack strategy to glean information of the key by estimating Alice's encoding via a quantum network, more specifically, the optimal-I network described in the earlier section and a security threshold is determined based on the errors in the key. Figure \ref{Fig4} illustrates Eve's attack on a bidirectional QKD protocol based on a $N=2$ quantum network.\\
\\

\subsection{Security Analysis}

\noindent
In this section, we shall commit to a simple security analysis of the bidirectional QKD protocol described above where we explore the use of the optimal-I network as an adversary's attack strategy against the protocol. Given Alice's choice of encoding is random and equiprobable, there should be no reason for Eve to prefer an optimal-I network based on the standard basis,  $\mathcal{T}_{\mathbb{U}}^{(2)}$, over $\mathcal{T}_{\mathbb{M}}^{(2)}$. Such a preference may in fact result in some non-natural noise which happens only in the cases where Alice uses one of the two operators. Hence we assume Eve would commit to either one of the two bases for her optimal-I network and the choice is also equiprobable. The optimal-I network, would provide a guess for Alice's choice of operator with the probabilities given in \ref{eq18}. In what follows, we explicitly provide the calculation for the case where Eve uses the $\mathcal{T}_\mathbb{U}^{(2)}$ network. The calculation for $\mathcal{T}_{\mathbb{M}}^{(2)}$ is completely analogous. 

It is instructive at this point to note some matters left wanting in this analysis. Mainly, we note that the attack here is one where Eve effectively interacts with the traveling qubits in each round of communication independently. The notion of coherent or collective attacks would simply be beyond the scope of this work. It is also worth noting that as the quantum network formalism is framed in the context of a POVM on the ancillae, or, operationally, Eve measuring her ancillae, her information gain is
quantified by Shannon entropy and this is done independently of Alice-Bob's public communications. Admittedly, this is a little discouraging as compared to a standard security analysis of a protocol which would measure Eve's Holevo information, quantified in terms of the von Neumann entropy. However, this prelimenary attempt in understanding such protocols with quantum networks may not be too far of the mark, as such protocols do not have any basis revelation process (like BB84) for which an eavesdropper could take advantage of before deciding on the best measurement to make.

\subsubsection{Mutual Information of Alice and Eve in Optimal-I Network} 

\noindent
Eve's uncertainty is based on her inference of Alice's encoding given her guess, particularly, the probability $\Pr(A_j|U_i)$ that Alice encoded using $A_j$ given her guess $U_i$. This is given by Bayes theorem and is explicitly calculated as,
\begin{align}
	\text{Pr}(A_1|U_i)  & = \begin{cases}
		\frac{4x^2+4xy+y^2}{5x^2+5xy+2y^2} ~\mbox{if}~ i=1
		 \\
		\frac{y^2}{x^2+xy+2y^2}  ~~~~\mbox{if}~ i \neq 1 
		\end{cases}, \label{eq30} \\
	\text{Pr}(A_2|U_i)  & =
	\begin{cases}
		\frac{x^2+xy+y^2}{5x^2+5xy+2y^2}  ~\mbox{if}~ i=1
		\\
		\frac{x^2+xy+y^2}{x^2+xy+2y^2} ~~~~\mbox{if}~ i \neq 1 
		\end{cases}. \label{eq31}
\end{align}
Eve's uncertainty quantified in terms of Shannon's entropy, $h(U_i)$ is easily calculated as,
\begin{eqnarray}\label{eq32}
	h(U_i) = -\sum_{j=1}^{2}\Pr(A_j|U_i) \log_2 \Pr(A_j|U_i).
\end{eqnarray}
As noted earlier, Eve should in principle randomly use either $\mathcal{T}_{\mathbb{U}}^{(2)}$ (with outcomes being the unitary operators $\{U_i\}$ as guesses), or $\mathcal{T}_{\mathbb{M}}^{(2)}$  (with outcomes being the unitary operators $\{M_k\}$) and her average uncertainty, $H_{AE}$, would be given as 
\begin{eqnarray} \label{eq33}
	H_{AE} = &~ \Big[\sum_{i=1}^{4} h(U_i) \Pr(U_i) + \sum_{j=1}^{4} h(M_k) \Pr(M_k)\Big],
\end{eqnarray}
where $\Pr(U_i)$ (or likewise for $ \Pr(M_k)$) is Eve's probability of getting $U_i$ on the whole, expressed as $\Pr(U_i) = [\sum_j\Pr(A_j|U_i)]/4$, where the factor of $1/4$ arises due to Eve's choice of selecting either $\mathcal{T}_{\mathbb{U}}^{(2)}$ or $\mathcal{T}_{\mathbb{M}}^{(2)}$, as well as Alice's independent equiprobable choice for encoding. The average information between Alice and Eve is given straightforwardly as $I_{AE}=1-H_{AE}$.

Given the constraint $x^2+xy+y^2=1$, we plot Eve's uncertainty for $i=1$ and $i\neq 1$ as functions of $y$, demonstrating how the case for $i\neq 1$ is preferable to her up to $y\approx 0.6$.
\begin{figure}[h]\label{Fig5}
	\centering
	\includegraphics[width=0.50\linewidth]{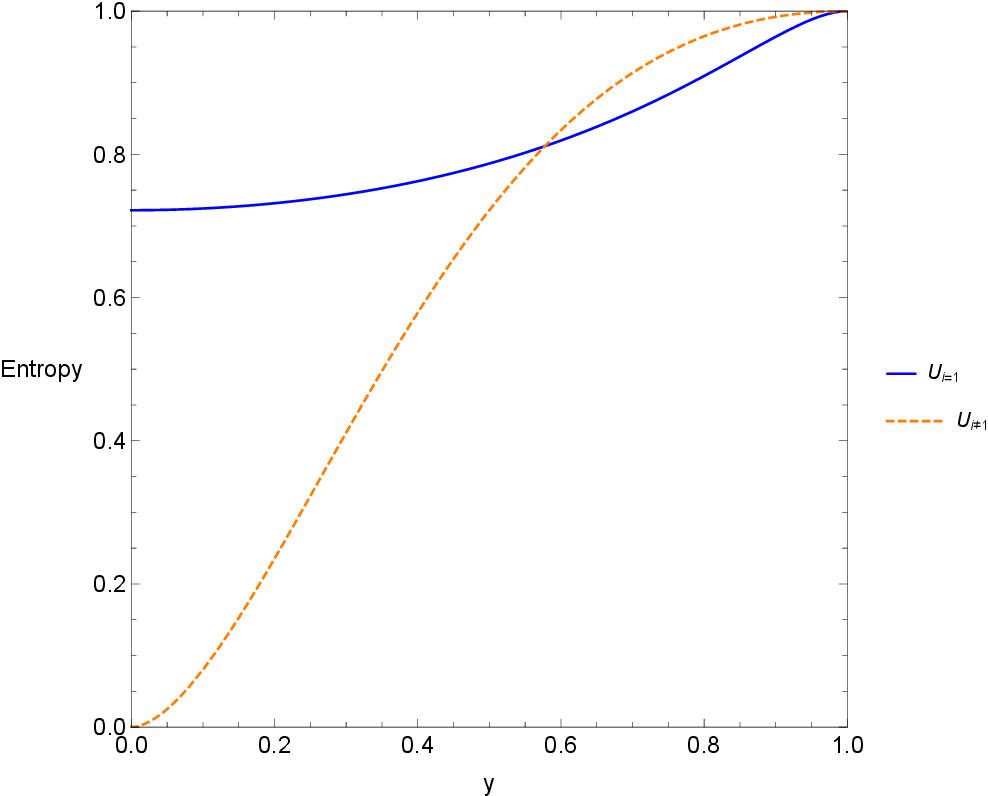}\\
	\caption{Eve's entropy according to Alice's encoding characterized by $y$.}
	%\label{fignorm}
\end{figure}\\
More precisely, for smaller values of $y$, where the `projective network' dominates, Eve gets more information in line with the decoding procedure where an outcome of the network coinciding with the actual encoding does not provide with a conclusive result. Note that in the case for  $i\neq 1$ and $x=1$, Eve's uncertainty is zero as she is effectively using the projective network and can make a conclusive inference of Alice's encoding. 
Conversely the minimum uncertainty in Eve's estimation of Alice's encoding is 0.73 for the case of $U_{i=1}$.

\subsubsection{Mutual Information of Alice and Bob in Optimal-I Network} 

Let Bob send a qubit, $|\phi \rangle$, chosen from his preferred basis for Alice's encoding, $A_{j}$. Unbeknownst to Alice and Bob, this qubit would undergo Eve's quantum channel (say, $\mathcal{T}_{\mathbb{U}}^{(2)}$) resulting in the final qubit, $| \phi_f (U{i})\rangle$, expressed as,
\begin{eqnarray}\label{34}
	| \phi_f (U_{i})\rangle =   y A_{j} | \phi \rangle + x \text{Tr}[A_{j}  U_{i}^\dagger] U_{i} | \phi \rangle .
\end{eqnarray}
This final qubit is then returned to Bob, who measures it in a chosen bases as described earlier. A post measured state of $| \psi_m \rangle$ occurs with the probability,
\begin{eqnarray}
	\Pr(|\psi_m\rangle |~ | \phi_f (U_{i})\rangle)& =|\langle \psi_m| \phi_f (U_{i})\rangle|^2 \label{eq35}\\
	& =  | y \langle \psi_m| A_{j} | \phi \rangle + x \langle \psi_m| U_{i} | \phi \rangle \text{Tr}[A_{j} U_{i}^\dagger] | ^2.\label{36}
\end{eqnarray} 
A conclusive but erroneous inference on Bob's side of Alice's operator can be detected by a sampling of the raw key and the average error due to Eve's interference is thus given by,
\begin{eqnarray}\label{eq37}
	E_{AB} = \frac{-2 + y^2 + y \sqrt{4 - 3 y^2}}{-6 + y^2 + y \sqrt{4 - 3 y^2}}.
\end{eqnarray}
Hence, with the average error between Alice and Bob leading to the mutual information between them, can be expressed as,
\begin{eqnarray}\label{eq38}
	I_{AB} = & ~ 1-h(E_{AB}).
\end{eqnarray}

\subsubsection{Security Threshold}

\noindent
Having established the mutual information between Alice and Eve, and Alice and Bob, we may determine the security threshold of the protocol given this attack strategy in terms of the intersection between the two. This is illustrated in the following graph.
\begin{figure}[h]\label{Fig6}
	\centering
	\includegraphics[width=0.6\linewidth]{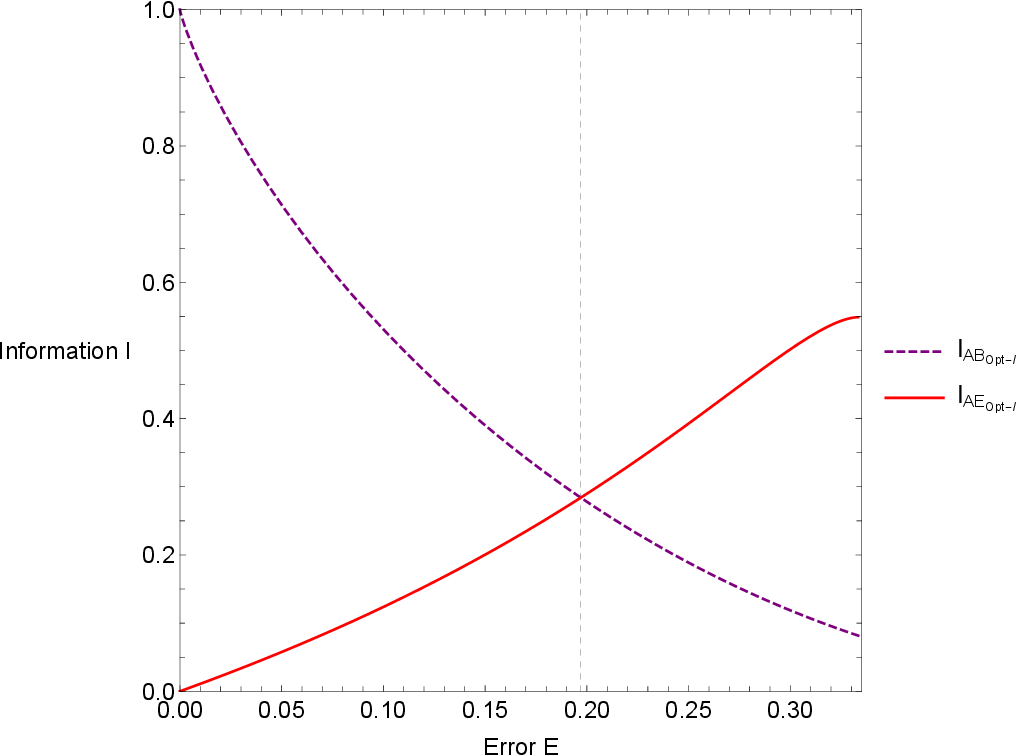}\\
	\caption{Mutual Information between Alice-Bob (dashed line) and Alice-Eve (solid line) versus the error rate E.}
	\label{fignorm}
\end{figure}\\
Given that Eve's information is quantified in terms of Shanon entropy, namely classical information, we may make use of the Csiszar-Korner theorem \cite{bib12} where $I_{AB} \geq I_{AE}$ ensures a secure QKD session. This is indicated in the graph where the error $0.197$ denotes the security threshold of the protocol for which beyond it, no secret key can be established betweeen Alice and Bob via error correction and privacy amplification. 

A quick comparison with the security threshold of $14.6\%$, for the BB84 protocol \cite{bib13} and that of a bidirectional QKD like LM05's $11.9\%$ \cite{bib6}, our current protocol, with a security threshold at $19.7\%$, demonstrates greater robustness than both BB84 and LM05. While this is not a definitive statement as it remains unclear if the considered quantum network represents the best strategy against such a protocol, we have believe strongly that  they may at least represent a close approximation to the most optimal strategies.

\section{Conclusion}

\noindent
In this work, we have described the structure of Quantum Networks in SU($2$) for finite outcomes. Despite its limitation in giving only a set of finite possible outcomes in stark contrast to the case of \cite{bib7} which allows for an outcome of any element of SU($2$), the information-disturbance tradeoff curves in both cases coincide. We conjecture this to be due to the notion that the standard basis and one which is mutually unbiased to it, which forms the basis of the quantum networks' outcomes, are the most different possible. Notably, in guessing a unitary from one finite basis using a quantum network that should provide an outcome mutually unbiased to it would result in a completely unbiased random guess. 

We then employ quantum networks as a nontrivial attack strategy against a  proposed bidirectional QKD protocol, inspired by the work of \cite{bib9}. In contrast to \cite{bib9}, which considers the use of unitary operators from a subspace of mutually unbiased unitary bases for Alice’s encoding, we consider unitary operators selected from two bases for the full $d^2$-dimensional space. The quantum network is introduced as an adversary, Eve, who attempts to guess the unitary operations performed by Alice while introducing as little disturbance as possible. It is revealed that the optimal-I network provides a security threshold of $19.70\%$, surpassing the security thresholds of the BB84 protocol ($14.6\%$) and LM05 ($11.9\%$).

Currently, there are no known formulations for quantum networks that are independent of Eve's measurement on the ancillary system. This aspect is of significant interest in the context of bidirectional QKD protocols to understand the security implications when Eve chooses to defer her measurement until after all classical discussions have been concluded between Alice and Bob. Further to that, it is also not known how a more comprehensive attack strategy akin to collective attack strategies related to encoding of quantum states used in protocols like BB84 may be constructed against bidirectional protocols. We envision the possibility of constructing a quantum network capable of testing multiple unitary operators, rather than testing one at a time. This involves utilizing a larger ancillary space to interact with various rounds of Bob's qubits and the best possible POVM that may measure the ancilla after all qubits’s tramnssion is complete. This more general quantum network may be indeed be an interesting avenue for future research.

\section*{Acknowledgement}

The authors, J S Shaari and N R S Abdul Salam, would like to express their gratitude to S Kamaruddin (IIUM) for helpful discussions and support. N R S Abdul Salam would like to thank the Ministry of Higher Education (Malaysia) for financial support through the Fundamental Research Grant Scheme For Research Acculturation of Early Career Researchers (FRGS-RACER) under grant number RACER19-043-0043 as well as the University's Research Management Centre for their assistance. The author S Mancini acknowledges financial support from ``PNRR MUR project PE0000023-NQSTI".

\end{document}